\documentclass{article}

\usepackage[letterpaper,top=2cm,bottom=2cm,left=3cm,right=3cm,marginparwidth=1.75cm]{geometry}

\usepackage{amsmath}
\usepackage{graphicx}
\usepackage{graphicx}
\usepackage{subcaption}
\usepackage[colorlinks=true, allcolors=blue]{hyperref}
\usepackage{orcidlink}
\usepackage{tabularx}
\usepackage[round,authoryear]{natbib}
\usepackage{xurl}

\title{A Framework for Teaching the Responsible and Effective Use of Generative AI in Statistics and Data Science}

\author{
    Vinny Davies\textsuperscript{1, *} \orcidlink{0000-0003-1896-8936},
    Craig Alexander\textsuperscript{1} \orcidlink{0000-0001-6734-747X},
    Jennifer Gaskell\textsuperscript{1} \orcidlink{0000-0001-9583-323X},  
    Matthew Parker\textsuperscript{1}, \\
    Jake Lever\textsuperscript{2} \orcidlink{0000-0001-8198-2939}
     \vspace{0.5cm} \\ 
    \textsuperscript{1}School of Mathematics and Statistics, University of Glasgow \\
    \textsuperscript{2}School of Computing Science, University of Glasgow \\
    \textsuperscript{*}Corresponding author email - Vinny.Davies@Glasgow.ac.uk \\
}
\date{}

\begin{document}
\maketitle

\begin{abstract}

The rapid rise of generative artificial intelligence (AI) following the public release of tools such as ChatGPT has begun to reshape how students engage with coding, statistical analysis, and written communication in higher education. Growing use of conversational AI systems creates opportunities for learning alongside challenges around reliability, transparency, privacy, and appropriate academic and professional practice, highlighting the need for structured education in their responsible use. This paper presents a framework for teaching the responsible and effective use of generative AI in statistics and data science. The framework is organised around four areas: Understanding, Development, Reporting, and Safety, covering conceptual foundations, practical interaction within analytical workflows, documentation and transparency, and issues including reliability, bias, privacy, and ethics. Its design draws on emerging work in AI literacy and established principles from statistics and data science education, and was further refined through workshop and conference activities and participant feedback. The framework has been implemented across multiple settings, including a massive open online course, short training sessions, and undergraduate and postgraduate skills-based teaching. We use these experiences to identify broader design considerations for educators, including the value of authentic analytical tasks, critical evaluation of generated outputs, integrated consideration of responsible practice, and flexibility as generative AI technologies continue to evolve.

\end{abstract}

\section{Introduction}

The public release of generative artificial intelligence (AI) tools such as ChatGPT in late 2022 rapidly changed how students engage with coding, statistical analysis, and written communication in higher education. Advances in neural network architectures, particularly the Transformer \citep{vaswani2017attention}, together with modern generative modelling approaches including diffusion models \citep{ho2020denoising}, have enabled systems capable of producing increasingly coherent text, code, images, and other forms of data through simple conversational interfaces. In statistics and data science, these systems can support tasks including coding, debugging, data exploration, interpretation, and communication, potentially changing both analytical workflows and expectations of computational tools.

Learners are already using generative AI to support coding, concept explanation, and written work \citep{pardos2023learning, freeman2025student, lepp2025does}, although the extent of adoption may sometimes be overestimated by educators \citep{lee2024impact}. Evidence on its educational impact is mixed and remains an active area of research. For some mathematics skills, ChatGPT-based support has been shown to produce learning gains comparable to those achieved with assistance written by human tutors \citep{pardos2024chatgpt}. However, improved performance while using generative AI does not necessarily translate into improved learning. In a large field experiment, unrestricted access to GPT-4 improved students' performance during mathematics practice but resulted in poorer performance once access to the tool was removed, while a more structured AI tutor largely mitigated this effect \citep{bastani2025generative}. Other experimental studies have reported lower long-term knowledge retention following ChatGPT-assisted study \citep{barcaui2025chatgpt}, as well as reduced cognitive engagement and poorer recall during AI-assisted essay writing \citep{kosmyna2025your}. Taken together, these studies illustrate both the potential benefits and risks of generative AI for learning, while also reflecting a rapidly developing evidence base in which study designs, contexts, and measures of learning vary considerably.

These findings highlight that the educational question is not simply whether generative AI should be used, but how it should be incorporated into learning \citep{UKCOTS26}. Given the widespread availability of these tools and their integration into commonly used software and online platforms, preventing their use entirely is unlikely to be practical even where this might be considered desirable. The increasing accessibility of generative AI therefore creates opportunities alongside challenges relating to reliability, transparency, academic practice, and appropriate use. Consequently, students need to be supported in using these tools critically, effectively, and responsibly \citep{nartey2024guiding,grove2025designing}.

Although research on AI literacy, responsible AI use, and AI safety provides useful insights, there is no single established structure for considering the effective and responsible use of generative AI within statistics and data science practice. Relevant perspectives address issues including understanding the capabilities and limitations of AI systems, privacy, accountability, ethics, governance, and potential harms \citep{chan2023comprehensive,hendrycks2025introduction}, but differ in their emphasis and intended context. Similarly, guidance in statistics and data science education emphasises conceptual understanding, authentic analytical problems, communication, responsible practice, and the effective use of technology within wider analytical workflows \citep{GAISE2016,deveaux2017curriculum,bile2021data}. However, these educational frameworks predate the widespread availability of generative AI and do not directly address its specific challenges and opportunities, leaving these considerations dispersed across different perspectives.

To address this gap, we developed a framework for teaching the effective and responsible use of generative AI in statistics and data science. The framework is organised into four areas: Understanding, Development, Reporting, and Safety. Understanding considers the concepts, capabilities, and limitations required for informed use. Development focuses on practical interaction with generative AI within analytical workflows. Reporting considers transparency, documentation, and reproducibility, while Safety addresses issues including reliability, bias, privacy, ethics, and organisational risk. The intention is not to prescribe a single model of generative AI use, but to provide a structure through which learners can critically consider when and how these tools should be incorporated into their work.

The framework was subsequently explored and refined through an initial workshop and later conference activities in which participants considered the use of generative AI for statistical and data science tasks \citep{ICOTS_Create}. Formative feedback from these activities informed revisions to the content and organisation. The resulting material has since been adapted for several educational settings, including a massive open online course (MOOC), short professional training sessions, and integration within undergraduate and postgraduate teaching. This paper describes the development of the framework, the rationale underlying its structure, and lessons emerging from its implementation, with the aim of providing transferable guidance for educators developing related material.

The remainder of the paper is organised as follows. Section~\ref{sec:courseplanning} describes the design considerations and formative development that informed the framework. Section~\ref{sec:content} presents the content and objectives of the four areas. Section~\ref{sec:courseformatassessment} describes the structured practical activities used to apply the framework, before considering different delivery formats and approaches to assessment. Finally, we discuss broader considerations for educators and future developments required as generative AI tools and modes of interaction continue to evolve.


\section{Course Planning}
\label{sec:courseplanning}

The framework was informed by three main sources: existing work on AI literacy and responsible AI use \citep[e.g.,][]{chan2023comprehensive,hendrycks2025introduction}, established guidance for statistics and data science education \citep[e.g.,][]{deveaux2017curriculum, bile2021data}, and our previous work considering how technological change is affecting statistics education \citep{UKCOTS26}. These ideas were explored and refined through an initial workshop in which participants considered how generative AI might be used within their own analytical practice \citep{ICOTS_Create}. The aim was not to prescribe a particular course or teaching format, but to develop a framework that could be adapted to different audiences and contexts. We then used this framework to develop an example of how its principles could be incorporated into teaching. This section describes the influences on the framework's design, the role of the initial workshop in its development, and the principles used to support its adaptation across different audiences and teaching formats.

\subsection{Motivation}

The existing literature provides useful guidance on generative AI literacy and responsible use, but does not offer a single established structure for considering its use within statistics and data science practice. Work on AI literacy emphasises understanding AI systems and evaluating their capabilities and limitations \citep{long2020ai}, while approaches in higher education highlight transparency, privacy, ethics, accountability, and institutional practice \citep{chan2023comprehensive}. Wider work on AI safety considers risks, potential harms, governance, and societal impact \citep{hendrycks2025introduction}. The design was also informed by established guidance in statistics and data science education, including an emphasis on conceptual understanding, active learning, realistic problems, communication, responsible practice, and effective use of technology \citep{GAISE2016,deveaux2017curriculum}. Together, these perspectives provided a basis for considering generative AI as part of the wider data analysis process, rather than as an isolated technical skill.

We used these ideas as a starting point rather than adopting an existing model directly, identifying four broad areas that captured the main practical questions associated with generative AI in statistics and data science: Understanding, Development, Reporting, and Safety. These formed the initial structure of the framework and were subsequently refined through the workshop and its associated activities. The framework was designed to remain flexible, recognising that requirements relating to privacy, reporting, assessment, and acceptable use vary across educational, research, and professional settings. The scope and content of the four areas, and their relationship to statistics and data science practice, are discussed in Section~\ref{sec:content}.

\subsection{Workshop and Initial Feedback}
\label{sec:initialfeedback}

The initial framework was explored through a one-day workshop designed to examine how generative AI might be used responsibly within data analysis practice and to provide formative feedback on the proposed approach \citep{ICOTS_Create}. Presentations introducing the main ideas were combined with structured reflection and discussion. Invited speakers from academic and professional backgrounds, including finance, law, AI, and higher education, shared examples of how generative AI was being used and considered across different sectors.

A central element of the workshop was the use of participants' own analytical problems. Participants considered where generative AI might contribute within an existing workflow and what issues its use might raise. An initial version of a structured worksheet supported this activity and encouraged reflection on different aspects of generative AI use. The framework and supporting materials were subsequently developed further following the workshop. The resulting four-area framework is described in Section~\ref{sec:content}, while the revised worksheet used in the current training materials is presented in Table~\ref{tab:worksheet} in Section~\ref{sec:courseformatassessment}.

Following the workshop, participants were invited to complete a short survey containing Likert-scale questions and open-response items. Although the number of survey responses was limited, these provided useful formative insights alongside verbal feedback gathered during the workshop and at subsequent conference presentations. This feedback informed the further development of the framework and supporting materials, rather than serving as a formal evaluation of a completed course.

Participants valued expert presentations, practical activities, and group discussion, which supported reflection on their own use of generative AI, although some suggested that additional time would allow more detailed reflection. They also emphasised the importance of practical guidance alongside conceptual understanding, reporting, and safety, favouring relatively simple, non-technical explanations of large language models. The observations informed the refinement of the framework and training materials, supporting an interactive, example-driven approach with structured reflective activities. The variation in requirements for reporting and safety across applications and sectors also reinforced the need for a consistent overall structure that could accommodate context-specific guidance.

\subsection{Audience and Adaptability}

A key design requirement was that the framework should be applicable to cohorts with differing levels of statistical, computational, and AI experience. It therefore focuses on the knowledge and considerations needed to use generative AI critically within data analysis, without assuming detailed knowledge of AI systems or the technical principles underlying machine learning. The intended learning outcomes (ILOs) should reflect the needs and prior experience of the target audience, with sufficient conceptual understanding to support critical and informed use of generative AI.

The framework was also designed to be adaptable across educational, research, and professional contexts, allowing the relative emphasis placed on individual topics to vary according to the audience. Learners with limited experience of generative AI may require greater emphasis on its capabilities and limitations, whereas professional audiences may place greater emphasis on organisational policy, privacy, reporting, or integration into existing workflows. This flexibility allows the framework to provide a consistent structure while accommodating different priorities through context-specific ILOs, examples, and guidance. Our implementations, described in Section~\ref{sec:courseformatassessment}, illustrate how the framework can be applied across different audiences and delivery formats.

\subsection{Practical Approach}
\label{sec:practicalapproach}

A central design principle was that learners should engage with generative AI through realistic data analysis problems rather than consider its use in isolation. This approach was reinforced by the initial workshop, where participants applied the emerging framework to problems from their own work (Section~\ref{sec:initialfeedback}; \citealp{ICOTS_Create}), and is also consistent with established recommendations in statistics education that emphasise active learning and the use of real data and authentic problems to develop statistical reasoning \citep{GAISE2016}. The practical activities therefore encourage learners to consider where generative AI might contribute within a wider analytical workflow and, equally importantly, where its use may be inappropriate or require additional scrutiny.

The practical activities use relatable examples to illustrate how generative AI might be incorporated into data analysis, while allowing learners to explore either their own analytical problems or a provided example. The structured worksheet supports this process and can be used with either type of problem, as described in Section~\ref{sec:courseformatassessment}. This approach encourages learners to consider where generative AI may be useful, where its use requires scrutiny, and how to verify outputs and report its use appropriately. The emphasis is on developing transferable judgement rather than proficiency with a particular system, allowing the framework to remain relevant as tools and working practices evolve.


\section{Framework Structure and Content}
\label{sec:content}

The design considerations described in Section~\ref{sec:courseplanning} resulted in four complementary areas: Understanding, Development, Reporting, and Safety. Understanding addresses the capabilities and limitations of generative AI; Development focuses on its practical use within analytical tasks and the role of human judgement; Reporting considers transparency and accountability; and Safety covers reliability, bias, privacy, ethics, and organisational risk. Although distinct, these areas are interconnected throughout a data analysis workflow. For example, using generative AI for coding involves understanding its limitations, verifying outputs, considering privacy, and reporting its contribution to the analysis. The framework provides a flexible structure rather than a fixed curriculum, allowing the emphasis placed on each area to vary across audiences, delivery formats, and institutional or professional contexts. Table~\ref{tab:ILOs} summarises our implementation, including the three topics within each area and their associated ILOs. The following subsections describe each area in more detail.

\begin{table}[t]
    \centering
    \begin{tabularx}{\textwidth}{l | p{2.4cm} | X}
        \textbf{Module} & \textbf{Topic} & \textbf{ILO} \\ \hline
        Understanding & Identifying Strengths & Explain how large language models generate text and the principles behind their outputs \vspace{0.2cm} \\
         & Understanding Limitations & Analyse the limitations and ethical challenges of generative AI, including bias, hallucinations, and lack of transparency \vspace{0.2cm} \\
         & Staying Safe & Demonstrate safe and responsible practices when using generative AI, including data privacy protection and avoiding plagiarism \vspace{0.2cm} \\ \hline
         
        Development & Learning ~ ~ ~ ~ Strategy & Develop an effective learning strategy that balances human effort and generative AI support, recognising its strengths and limitations \vspace{0.2cm} \\
         & Development Strategy & Understand how to apply generative AI responsibly within projects, including data exploration, development, evaluation, and documentation of AI use \vspace{0.2cm} \\
         & Effective ~ ~ ~ ~ Usage & Explore effective day-to-day use of generative AI, understanding when AI is appropriate and when human judgement is required \vspace{0.2cm} \\ \hline
         
        Reporting & Generative Process & Analyse how generative AI has been used in work contexts and determine when its use should be declared \vspace{0.2cm} \\
         & Regulations \& Reporting & Evaluate reporting practices and relevant regulations for responsible use of generative AI across different scenarios \vspace{0.2cm} \\
         & Ethical ~ ~ ~ ~ Considerations & Assess ethical considerations of generative AI in projects, understanding their impact on work and organisational practice \vspace{0.2cm} \\ \hline
         
        Safety & Education \& Understanding & Develop strategies for educating and supporting staff in the use of generative AI within an organisation \vspace{0.2cm} \\
         & Risk ~ ~ ~ ~ ~ ~ ~ ~ Management & Analyse risks associated with adopting generative AI and identify approaches to mitigate them \vspace{0.2cm} \\
         & Processes \& Documentation & Design processes and documentation to support responsible and consistent use of generative AI in organisational practice \vspace{0.2cm} \\
    \end{tabularx}
    \caption{Structure of the framework showing the topics within each area and the associated ILOs}
    \label{tab:ILOs}
\end{table}

\subsection{Understanding}
\label{sec:understanding}

The Understanding component addresses the need identified during development of the framework for learners to have sufficient conceptual knowledge to evaluate generative AI critically, without requiring detailed technical expertise. Accordingly, the depth and technical content are adapted to the audience, with our implementation providing a high-level account of how large language models operate that focuses on a useful conceptual model rather than the underlying theory or architecture. Neural network structures and the mechanics of text processing are omitted to keep the material accessible and focused on practical understanding. Instead, large language models are introduced as systems that predict likely continuations of a sequence, providing a basis for exploring their capabilities and limitations.

Topic 1, Identifying Strengths, explores how generative AI systems generate text and where their information originates. To make the predictive generation more understandable to learners, we use an invented `alien language' composed of emojis, asking learners to predict the next symbol without knowing its meaning. This illustrates how patterns can be identified without semantic understanding and provides an accessible introduction to the black box nature of these systems. We then explore how probability distributions influence output selection, explaining why responses can vary and sometimes appear creative or unexpected. We then turn to the origins of model outputs, outlining how models are trained on large text corpora before introducing retrieval augmented generation (RAG). This illustrates how connecting models to external sources can influence and constrain the information available in their responses.

Topic 2, Understanding Limitations, examines the implications of probabilistic generation and training data for the reliability of model outputs. We introduce hallucinations as confident but incorrect outputs, emphasising that generated responses are not equivalent to information retrieved from a verified knowledge base. Rather than cataloguing examples that may quickly become outdated, we focus on the mechanisms that can give rise to unreliable outputs. We also examine how biases can enter models through their training data, including inaccuracies and socially embedded prejudices relating to characteristics such as gender and ethnicity. These examples highlight how the processes that enable useful outputs can also contribute to limitations. Finally, we consider the environmental implications of generative AI, including the energy consumption associated with training and deploying large-scale systems.

Topic 3, Staying Safe, translates this understanding into considerations for responsible use. We begin with data privacy, explaining why personal or sensitive information requires particular care when generative AI systems are used and why users need to understand the policies governing the tools available to them. We then address verification, considering how and why outputs should be critically evaluated and cross checked before being relied upon. From there, we consider the cognitive implications of regular use, including risks associated with over-reliance and cognitive offloading \citep{kosmyna2025your,zhai2024effects}, alongside the potential benefits for learning when tools are used reflectively. Finally, we examine plagiarism and intellectual property, including the possibility of unintentionally reproducing the work of others, while reserving more detailed consideration of reporting and organisational responsibilities for later components.

Taken together, the Understanding module equips learners with a practical conceptual model of generative AI, enabling them to recognise its capabilities, limitations and implications for responsible use. This knowledge is intended to support informed judgements about when generated outputs may be useful, when they require further scrutiny, and when the use of generative AI may be inappropriate. These judgements provide the foundation for the Development module, where learners apply them more directly within statistics and data science workflows.

\subsection{Development}
\label{sec:development}

The Development component translates these principles into practical use within statistics and data science. Its emphasis reflects the framework's broader aim of developing judgement rather than simply proficiency with a particular tool or technique. The material was developed before approaches such as ``vibe coding'' \citep{sarkar2025vibe}, in which development is increasingly delegated to AI agents, became prominent, illustrating how quickly the ways in which generative AI can support development may change. Rather than attempting to anticipate or prescribe specific approaches, the framework focuses on transferable principles for incorporating generative AI into analytical workflows. Although the specific technologies and activities can vary according to the audience and application, the central aim is to support a structured and reflective approach to AI-assisted development, in which learners consider both where generative AI can contribute and where human understanding and judgement remain necessary.

Topic 1, Learning Strategy, considers the learning process that precedes the use of generative AI for development tasks. Before relying on AI to assist with coding or modelling, learners are encouraged to consider the task itself and the knowledge required to complete it. The focus is on constructing a plan that connects their current knowledge to the intended outcome. Generative AI may support this process, for example by helping to identify relevant resources or clarify unfamiliar concepts, but learners are encouraged to decide which aspects of the process are best undertaken independently and where AI can provide useful support. Examples from statistics and data science illustrate how apparent improvements in task completion do not necessarily correspond to improved understanding. These considerations inform the learning plan, which learners revisit as they progress through the module.

Topic 2, Development Strategy, focuses on incorporating generative AI into the development process. We begin by revisiting questions of appropriate and safe use, linking back to the issues discussed in Section~\ref{sec:understanding}. For example, learners consider whether uploading data to a generative AI tool is appropriate and permitted within their working context. The discussion then turns to identifying which stages of a development task are suitable for AI assistance and which should remain primarily the responsibility of the user. Practical considerations include documenting AI contributions, refining requests through iterative interaction, and testing generated outputs. Rather than focusing on detailed prompt-engineering techniques, the emphasis is on maintaining a transparent and thoughtful development process.

Topic 3, Effective Usage, builds on these principles by considering practical approaches to interacting with generative AI tools. The discussion remains general so that the guidance continues to be relevant as tools evolve. We introduce prompt engineering in a broad sense, examining how additional context, specification of the intended audience, examples, and clearer descriptions of the task can influence the usefulness of generated outputs. Learners are encouraged to inspect, test, and refine outputs iteratively rather than accept them uncritically. We also consider how the increasing integration of generative AI into development environments may change these interactions, reinforcing the value of principles that remain applicable across different interfaces and systems.

Overall, the Development module encourages learners to treat generative AI as one component of a wider analytical workflow rather than as a substitute for the workflow itself. Planning, verification, reflection, and human judgement remain central to this approach. These principles lead naturally to the Reporting module, which considers how the contribution of generative AI should be made visible to others.

\subsection{Reporting}
\label{sec:reporting}

The Reporting component reflects the design principles of transparency and accountability by considering how the use of generative AI should be documented and communicated within academic, professional, and technical work. Appropriate practice varies according to the type of work, the organisational context, and the extent of the AI contribution, so the framework does not prescribe a single reporting standard. Instead, learners are encouraged to consider what others would need to know to understand the contribution made by generative AI, how the resulting work was checked, and what external requirements may apply.

Topic 1, Generative Process, begins by asking learners to reflect on how generative AI has contributed to their work and whether that contribution should be declared. The discussion distinguishes relatively minor uses, such as proofreading or debugging, from more substantial involvement, including generating code, contributing to an analytical approach, or producing larger parts of a report or project. Learners also consider how they interacted with the tool, including whether prompts were refined iteratively and how different approaches influenced the outputs. Particular emphasis is placed on verification, for example by checking references, cross-checking information against trusted sources, or testing generated code and models. This reflection helps learners consider both whether AI use should be declared and the level of detail that may be appropriate.

Topic 2, Regulations and Reporting, considers how such contributions can be documented in practice and how reporting is shaped by internal and external requirements. Relevant information may include the tool used, its contribution to the work, prompts or outputs, subsequent edits, and the methods used to verify the resulting material. The appropriate level of detail depends in part on the extent of the contribution: relatively minor uses may require only a brief acknowledgement, whereas more substantial involvement may warrant a dedicated description in the methodology or project documentation. The topic also introduces organisational policies, publisher requirements, data protection, and other external regulations that may restrict or shape the use of generative AI. Reporting is therefore presented not only as a means of promoting transparency, but also as a way to support accountability and reproducibility.

Topic 3, Ethical Considerations, revisits issues introduced earlier in the framework and examines their implications for individual projects and organisations. These include bias, environmental impact, and intellectual property or plagiarism. Learners are encouraged to consider how biased outputs might influence decisions or affect different groups, how the environmental costs of generative AI relate to wider sustainability goals, and how generated material may raise intellectual property concerns. Reporting is considered alongside these issues because transparent documentation makes the use of generative AI more visible and open to scrutiny. The emphasis is therefore on recognising that responsible use extends beyond compliance with formal requirements to encompass the broader consequences of using these tools.

Overall, the Reporting module progresses from reflecting on the contribution of generative AI, through documenting and communicating that contribution, to considering the wider regulatory and ethical implications. It emphasises transparency and accountability while recognising that appropriate practice depends on context. The Safety module extends this perspective from individual projects to the organisational structures and processes needed to support responsible use.

\subsection{Safety}
\label{sec:safety}

The Safety component extends the framework from the individual learner or project to the wider organisational context. This reflects the design principle that responsible use is shaped not only by individual behaviour, but also by the policies, processes, training, and support available within the environment in which generative AI is used. As risks and responsibilities vary between organisations, the component provides a structure for considering these issues rather than prescribing a fixed set of requirements. Its three topics address staff education, risk identification and mitigation, and the processes needed to embed responsible practice within organisational workflows.

Topic 1, Education and Understanding, focuses on ensuring that staff have sufficient knowledge to use generative AI effectively and responsibly. This includes awareness of its strengths and limitations, data privacy and security, approaches to testing and verification, and the range of available tools. The topic also considers whether education should be delivered across an organisation or tailored to particular roles, recognising that generative AI may be used differently across areas of work. Existing training may need to be adapted, for example by incorporating generative AI into data protection guidance, while new forms of interaction may require additional support. Practical approaches include assessing existing levels of knowledge and use, providing regular training and guidance, restricting access to approved tools where appropriate, and identifying individuals who can provide further support.

Topic 2, Risk Management, considers the risks associated with introducing generative AI into organisational workflows and how these can be mitigated. Some risks, such as unreliable generated information, arise from the characteristics of generative AI itself, while others, such as data breaches, depend strongly on how the tools are used. Their likelihood and potential consequences also vary according to the task and context. Examples include data breaches, inaccurate claims, and copyright or intellectual property infringements, which may have legal, financial, or reputational consequences. Risk mitigation strategies include human oversight, verification and testing, clear internal policies, restrictions on the use of particular tools or data, staff training, and formal risk assessment where appropriate. The topic also emphasises the need to review these measures as technologies and patterns of use evolve.

Topic 3, Processes and Documentation, considers how responsible practice can be embedded within organisational workflows. We begin by asking who should be involved in developing relevant processes, including management, those carrying out the work, and individuals with appropriate technical or professional expertise. The discussion then considers useful forms of documentation, such as descriptions of the purpose of generative AI use, records of significant AI contributions, reporting procedures, lists of approved tools, and records of how outputs were verified. Different processes may be appropriate for different contexts, including coding, research, administrative activity, and the production of formal reports. Particular emphasis is placed on proportionality: overly detailed procedures may become burdensome, while insufficient documentation may leave important risks unmanaged. Risk assessments provide one example of a structured approach, allowing the source and nature of a risk, its likelihood and severity, potential consequences, mitigation strategies, and residual risk to be considered explicitly.

Overall, the Safety module considers how organisations can support responsible use through staff education, proportionate risk management, and appropriate processes and documentation. Together, these elements reinforce the framework's central principle that effective and responsible use depends not only on the capabilities of generative AI systems, but also on the judgement of those using them and the organisational environment in which that use takes place.


\begin{table}[p]
    \centering
    \begin{tabularx}{\textwidth}{p{2.5cm} | X}
        \textbf{Section} & \textbf{Discussion Prompt} \\ \hline
        
        \textbf{Understanding} & \vspace{0.5cm} \\
        Identifying Strengths & What are the relevant strengths of Generative AI? How do these align with the various tasks and skills your project needs? Where do Generative AI's strengths overlap with your project's demands, and where might it therefore make sense to consider its use? \vspace{0.2cm} \\
        Understanding Limitations & What are the relevant limitations of Generative AI? Considering your project's specific tasks and skills, how do these demands intersect with its weaknesses? Where might it, therefore, be prudent to consider avoiding the use of Generative AI to mitigate potential issues? \vspace{0.2cm} \\
        Staying Safe & What security considerations are relevant when using Generative AI for this project? Are there any privacy, confidentiality, or regulatory issues that must be addressed? What is the intended use of the generated content, and how will you ensure it performs as expected? Are there any ethical considerations that should be taken into account? \vspace{0.2cm} \\ \hline
        
        \textbf{Development} & \vspace{0.5cm} \\ 
        Learning ~ ~ ~ ~ Strategy & What do you need to learn to be able to complete this project? How will you link what you need to learn with what you already know? How will you verify what you have learnt from Generative AI? \vspace{0.2cm} \\
        Development Strategy & What tasks are you going to use Generative AI for? How are you going to verify that what you produce is correct or meets the required standard? How will the parts you complete using Generative AI interact with the parts you do yourself?\vspace{0.2cm} \\
        Effective Usage & How are you going to structure the tasks that you ask the Generative AI tool to do? Will you break the tasks down into logical parts, or ask it to do everything and then refine the output? Will you use the tools primarily for generation, refinement, or both? \vspace{0.2cm} \\ \hline
        
        \textbf{Reporting} & \vspace{0.5cm} \\
        Generative ~ ~ ~ Process & When using Generative AI tools in your project, should their involvement be declared? Is it necessary to detail the generation and verification processes? How would you check its integration within the wider code and documentation?\vspace{0.2cm} \\
        Regulations \& Reporting & What internal regulations are in place concerning the use of Generative AI? Has this matter been discussed at management level? Are there any specific reporting requirements related to its application? Furthermore, which external regulations or reporting obligations might be affected by the adoption of Generative AI? \vspace{0.2cm} \\
        Ethical ~ ~ ~ ~ ~ ~ Considerations & Generative AI is often prone to inherent biases and potential for unfairness. Could any of these issues inadvertently manifest within your project? Is your project truly as equitable as it could be for all users, ensuring everyone is considered to the same extent you would ordinarily apply? Furthermore, does it unfairly utilise existing intellectual property, going beyond what is currently accepted practice for Generative AI tools? \vspace{0.2cm} \\ \hline
        
        \textbf{Safety} & \vspace{0.5cm} \\
        Education \& Understanding & Do the team members adequately comprehend Generative AI tools, enabling both their practical application and an awareness of potential implications? What strategies can be implemented to ensure their skills are brought up to the required standard? \vspace{0.2cm} \\
        Risk ~ ~ ~ ~ ~ ~ ~ Management & What inherent risks exist for the organisation and its particular field? How do these considerations shape the application and record-keeping practices for Generative AI? \vspace{0.2cm} \\
        Processes \& Documentation & What processes and documentation should be implemented? What level of detail and scrutiny is needed to ensure diligence without unnecessary effort? \vspace{0.2cm} \\
    \end{tabularx}
    \caption{Discussion prompts used across our different implementations of the framework}
    \label{tab:worksheet}
\end{table}

\section{Delivery Formats and Assessment}
\label{sec:courseformatassessment}

The framework is designed to support different teaching formats rather than prescribe a single course structure. At the centre of our implementations are structured practical activities that encourage learners to apply the framework to realistic analytical problems. These activities can be used in standalone training sessions, online courses, or existing statistics, data science, and research-skills teaching. The depth of each topic, the balance between presentation and discussion, and the approach to assessment can be adapted to suit the audience, available teaching time, and institutional or professional context. In this section, we first describe the structured practical activities and worksheet used to support application of the framework. We then describe three implementations: a MOOC, standalone training sessions, and integration within existing skills-based teaching.

\subsection{Structured Practical Activities}
\label{sec:practicalactivities}

A structured worksheet supports learners in applying the framework to an analytical problem. Learners can select a problem from their own work or studies and revisit it as they progress through Understanding, Development, Reporting, and Safety. Rather than treating generative AI as something that should simply be ``used'' or ``not used'', the activity encourages consideration of where it might contribute within a wider workflow, what limitations or risks may arise, and what forms of verification, reporting, or organisational support may be required. An initial version of the worksheet was used in the workshop described in Section~\ref{sec:initialfeedback}, with the worksheet subsequently refined alongside the framework and teaching materials. The current discussion prompts are shown in Table~\ref{tab:worksheet}.

Alongside learner-selected problems, we use a common worked example involving a public-sector analysis of sensitive patient data and the production of a report for public dissemination:

\begin{quote}
\emph{Imagine you work in the public sector and have been tasked with analysing a dataset containing information on a group of patients, including whether they have diabetes and other sensitive clinical details. You have been asked to produce a report based on your findings, which will be published on an official website and made publicly available.}
\end{quote}

The example is revisited throughout the framework, allowing learners to explore appropriate use, verification of generated outputs, reporting, privacy, and organisational risk within a consistent scenario. It is particularly useful in settings with open or semi-open discussion forums, such as the MOOC described in Section~\ref{sec:MOOC}, where discussing examples from learners' own work could risk disclosing sensitive or confidential information. The shared scenario provides a consistent basis for discussion without requiring learners to disclose details of their own work, while learner-selected problems allow them to consider how the same principles apply in their own analytical contexts. These activities can also support assessment, with short formative questions testing key concepts and discussion, reflection, and worksheet activities exploring more context-dependent decisions, including how generative AI outputs should be verified, documented, and managed.

\subsection{Generative AI for Data Science MOOC}
\label{sec:MOOC}

One implementation of the framework is a MOOC delivered on the Coursera platform, organised into four modules based on the content described in Section~\ref{sec:content}. Topics are sometimes divided into shorter segments, as shorter videos and learning units have been associated with improved learner engagement in online courses \citep{guo2014video}. The course combines videos and written material, with some overlap between the formats to support different ways of engaging with the content. Each topic includes a multiple-choice formative assessment, while each module concludes with a multiple-choice summative assessment. Short introductory and closing segments are also included within each module, alongside initial material introducing the course aims and the staff involved in its development and delivery. The course is available at \url{https://www.coursera.org/learn/gen-ai-for-data-science}.

The common worked example introduced in Section~\ref{sec:practicalactivities} is used throughout the MOOC to illustrate how the framework can be applied to a practical analytical workflow. Each module includes a video exploring the relevant ideas through this example, helping learners connect conceptual material with practical situations. Concrete and relatable examples can support learning by helping learners connect abstract concepts with practical situations \citep{umutlu2022leveraging}. The course forums provide further opportunities for reflection and discussion, using prompts based on the activities in Table~\ref{tab:worksheet}. Learners can consider either the supplied example or an appropriate problem of their own, with the shared example particularly useful in this setting because discussions are visible to other participants and may be unsuitable for detailed discussion of confidential or sensitive projects.

\subsection{Training Sessions}
\label{sec:training}

The framework has also been delivered through standalone training sessions lasting two to three hours. These cover the same core material in a condensed format, using a structured presentation with breaks for discussion and practical activities based on the worksheet in Table~\ref{tab:worksheet}. Participants are encouraged to select a problem from their own work or studies and revisit it throughout the session, helping them consider the role of generative AI within a specific workflow. Where participants lack a suitable example or confidentiality prevents discussion of their own work, the common worked example described in Section~\ref{sec:practicalactivities} can be used instead. The depth of individual topics, choice of examples, and balance between presentation and discussion can be adapted to the audience and time available. Professional sessions may place greater emphasis on organisational policy, risk, and reporting, while sessions aimed at students may devote more time to learning strategies, verification, and the relationship between generative AI use and independent analytical development.

Printed worksheets are provided for participants to annotate and retain for future reference. The materials are also available at \url{https://github.com/vinnydavies/generative-ai-for-data-science/}, allowing other educators to adopt or adapt the approach for their own teaching. The sessions have been delivered in a range of contexts, and further targeted training is planned ahead of undergraduate and postgraduate dissertation projects, where students are likely to encounter decisions about appropriate generative AI use during substantial independent analytical work.

\subsection{Skills Course Integration}
\label{sec:skillsintegration}

The framework can be incorporated into existing undergraduate and postgraduate skills-based teaching rather than delivered as a separate course. Its four areas need not be taught consecutively, with individual components introduced where they are most relevant to the wider curriculum. For example, Understanding and Development can be integrated into coding, statistical analysis, or data-science activities, while Reporting can be incorporated into teaching on reproducibility, research communication, and academic or professional practice. Safety can be linked to existing material on ethics, data protection, governance, and responsible research. The worksheet in Table~\ref{tab:worksheet} can provide continuity across these activities, with learners returning to the same problem at different stages of a course or project and reconsidering their decisions as they encounter new material.

This approach may be particularly useful for educators who lack the time or resources to develop a dedicated generative AI course, as relevant elements of the framework can be incorporated into established activities according to the programme and learners' prior experience. In our own teaching, we plan to adopt this approach within skills courses supporting undergraduate and postgraduate project work, introducing relevant material before students begin substantial independent analytical work and revisiting it as their projects progress. This gives learners repeated opportunities to consider when generative AI supports their work, how outputs should be scrutinised, and when greater independence or additional safeguards may be required.


\section{Discussion}

The work presented in this paper describes the development of a framework to support effective and responsible engagement with generative AI in statistics and data science. In response to the rapid adoption and increasing accessibility of these tools, the framework seeks to move beyond approaches that either encourage unrestricted adoption or focus primarily on preventing their use. Instead, it aims to help learners integrate generative AI into analytical workflows in a critical, reflective, and transparent way. The four areas of Understanding, Development, Reporting, and Safety bring together conceptual knowledge, practical application, transparency and accountability, and responsible use within a common structure that can be adapted to different educational and professional contexts. This approach aligns with wider work on AI literacy and responsible AI use, which emphasises the importance of understanding AI capabilities and limitations alongside the implications of their use \citep[e.g.,][]{chan2023comprehensive,hendrycks2025introduction}.

The development process outlined in Section~\ref{sec:courseplanning} highlights several principles that may inform the design of similar educational material. Early formative engagement through the workshop helped refine the framework's structure and reinforced the importance of accounting for variation in learners' prior knowledge. It also supported an approach grounded in realistic analytical problems rather than treating generative AI as an isolated technical skill. This is consistent with established recommendations in statistics and data science education that emphasise active learning, authentic problems, conceptual understanding, and the effective use of technology \citep{GAISE2016,deveaux2017curriculum, bile2021data}. More broadly, the framework places emphasis not simply on learning how to use generative AI, but on developing the judgement needed to decide when its use is appropriate and how its outputs should be evaluated.

The four areas described in Section~\ref{sec:content} are deliberately interconnected and can be revisited throughout a data analysis workflow rather than treated as independent topics. Their implementation is also intended to remain flexible. The topics and intended learning outcomes presented in Table~\ref{tab:ILOs} represent our implementation of the framework rather than a fixed curriculum, allowing the emphasis placed on individual areas to vary according to learner experience, application, and institutional or professional requirements. The practical activities described in Section~\ref{sec:practicalactivities} similarly combine a common worked example with learner-selected analytical problems. This allows general principles to be demonstrated consistently while encouraging learners to consider how they apply within their own work. Concrete and relatable examples can help learners connect more abstract concepts to practical situations \citep{umutlu2022leveraging}, while a shared fictional example is particularly valuable where discussion takes place in open or semi-open settings and disclosure of real project information may be inappropriate.

The implementation examples in Section~\ref{sec:courseformatassessment} demonstrate how the framework can be delivered in different settings. We have used it to develop a MOOC and standalone training sessions, and plan to integrate its components into existing undergraduate and postgraduate skills-based teaching. A dedicated course will not be feasible or necessary in every setting; the framework can also inform shorter training sessions or be incorporated selectively into teaching on statistics, data science, research methods, and professional skills. The depth of coverage, balance between the four areas, use of the worksheet, and approach to assessment can be adapted to the time available, learners' experience, and local requirements.

Several broader recommendations emerge from this work. Teaching about generative AI should be grounded in realistic analytical tasks rather than focusing primarily on particular tools or prompting techniques, especially as technologies and interfaces continue to evolve. Conceptual explanations should provide sufficient understanding to support critical use without introducing unnecessary technical barriers. Verification, transparency, privacy, and responsible practice should also be integrated throughout teaching rather than treated as separate concerns. Finally, educators should distinguish between improving task performance and supporting learning. Emerging evidence suggests that excessive reliance on generative AI may encourage cognitive offloading or reduce engagement with some of the reasoning processes that learners would otherwise undertake \citep{kosmyna2025your,zhai2024effects}. Teaching should therefore create opportunities for learners to benefit from these tools while continuing to develop, practise, and demonstrate their own understanding and judgement.

The framework should be viewed as a starting point for further development rather than a complete or permanent curriculum. Feedback gathered during its initial development was formative and intended to refine the structure and teaching approach, rather than provide a formal evaluation of learning outcomes. Verbal feedback from the initial workshop and subsequent presentations was positive, but only limited formal survey data were collected. The framework's effectiveness in supporting learning and responsible practice therefore remains to be established through more systematic evaluation. Future work will examine its implementation across different cohorts and contexts, alongside continued refinement of the activities, examples, and assessment approaches. Such evaluation will be important given that the educational effects of generative AI are likely to vary according to the task, context, learner, and manner of use.

A further area for development is the increasing integration of generative AI into programming environments, including AI coding assistants embedded within editors such as Visual Studio Code and tools supporting R-based workflows. These approaches were less established when the original teaching materials were developed and are not yet considered in detail. Future implementations will need to address increasingly autonomous approaches to code generation, including practices sometimes described as ``vibe coding'', in which substantial parts of software development are undertaken through natural-language interaction \citep{sarkar2025vibe}. While these approaches may accelerate development and lower barriers to programming, they also reinforce the importance of understanding, testing, validating, and taking responsibility for generated code. The four-area structure provides a basis for incorporating such developments while maintaining an emphasis on critical, transparent, and responsible use.

\section*{Acknowledgements}

Data collection protocols were reviewed and approved by the University of Glasgow Ethics Committee, and the authors thank the study and workshop participants for their contributions. The authors also acknowledge funding from the University of Glasgow Centre for Data Science and AI to support the workshop and funding from the University of Glasgow Learning Innovation Support Unit to support development of the course. The authors acknowledge the use of generative AI tools to assist with editing and improving the clarity of the manuscript.



\bibliographystyle{abbrvnat}
\bibliography{main}

\end{document}